 \documentclass[reprint,prb,floats,showpacs,onecolum,twocolumn]{revtex4-1}

\usepackage[english]{babel}
\usepackage{amsmath}
\usepackage{amssymb, comment}
\usepackage[]{graphicx}
\usepackage{times}
\usepackage{bm}
\usepackage{units}
\usepackage{color}

\renewcommand{\eqref}[1]{Eq. (\ref{#1})}

\newcommand{\PL}{\textnormal{PL}}

\begin{document}
\title{Molecule signatures in photoluminescence spectra of transition metal dichalcogenides}

\author{Maja Feierabend$^1$, Gunnar Bergh\"auser$^1$, Malte Selig$^{1,2}$,  Samuel Brem$^1$, Timur Shegai$^1$, Siegfried Eigler$^3$ and Ermin Malic$^1$}
\address{$^1$Chalmers University of Technology, Department of 
Physics, 412 96 Gothenburg, Sweden}
\address{$^2$Institut f\"ur Theoretische Physik, Technische Universit\"at Berlin, 10623 Berlin, Germany}
\address{$^3$Institiut f\"ur Chemie und Biochemie, Freie Universit\"at Berlin, 14195 Berlin, Germany}

\begin{abstract}
Monolayer transition metal dichalcogenides (TMDs) show an optimal surface-to-volume ratio and are thus promising candidates for novel molecule sensor devices. It was recently predicted that a certain class of molecules exhibiting a large dipole moment can be detected through the activation of optically inaccessible (dark) excitonic states in absorption spectra of tungsten-based TMDs. In this work, we investigate the molecule signatures in photoluminescence spectra in dependence of a number of different experimentally accessible quantities, such as excitation density, temperature as well as molecular characteristics including the dipole moment and its orientation, molecule-TMD distance, molecular coverage and distribution. We show that under certain optimal conditions, even  room temperature detection of molecules can be achieved.
\end{abstract}

\maketitle
\section{Introduction}
Transition metal dichalcogenides (TMDs) present a promising class of nanomaterials with a direct band gap, efficient electron-light coupling, and strong Coulomb interaction
\cite{THeinz, He2014, splendiani2010emerging, arora2015excitonic,gunnar_nano}.
 The latter gives rise to a variety of tightly bound excitons, which determine the optical response of TMDs 
\cite{Chernikov2014, gunnar_prb,steinhoff2014influence}. 
As atomically thin materials, they  show an optimal surface-to-volume ratio and are therefore very sensitive to changes in their surroundings 
\cite{raja2017coulomb, schmidt2016reversible, conley2013bandgap}.
 A consequence is that one can tailor the optical fingerprint of these materials through external molecules 
\cite{voiry2015covalent,yuan2014establishing,yong2014ws,hayamizu2016bioelectronic}. Applying non-covalent functionalization 
\cite{Hirsch2005}, the electronic band structure remains to a large extent unaltered, while optical properties significantly change.\\
In our previous work, we have proposed a new sensing mechanism for molecules based on the activation of dark excitonic states in monolayer TMDs \cite{maja_sensor, ermin_review}. 
These dark states can lie energetically below the bright ones but are not directly accessible by light as they are either spin or momentum forbidden \cite{hoegeleMono,hoegeleBi,zhang2015experimental}.
We have shown that in the presence of molecules the absorption spectra exhibits an additional peak appearing at the position of the optically inaccessible $K\Lambda$ exciton.  Unfortunately, the peak is only visible considerably below room temperature due to the significant broadening of excitonic transitions \cite{selig2016excitonic}. The aim of this work is to investigate the possibility to achieve room-temperature sensing of molecules. To reach this goal we  investigate  photoluminescence (PL) spectra, since in contrast to absorption they are not characterized by a large background signal and thus show a better signal-to-noise ratio. Another important advantage of PL is its strong dependence on the number of excited excitons. After the process of thermalization, the excitons  mainly occupy the energetically lowest dark $K\Lambda$ state \cite{selig2017dark}. As a result, the additional dark exciton peak is expected to be very large compared to the bright peak that otherwise strongly dominates absorption spectra. This work reveals the optimal conditions for the maximal visibility of dark excitons in PL spectra, which presents an important step towards a possible technological application of TMDs as molecular sensors.

\begin{figure}[t]
  \begin{center}
\includegraphics[width=\linewidth]{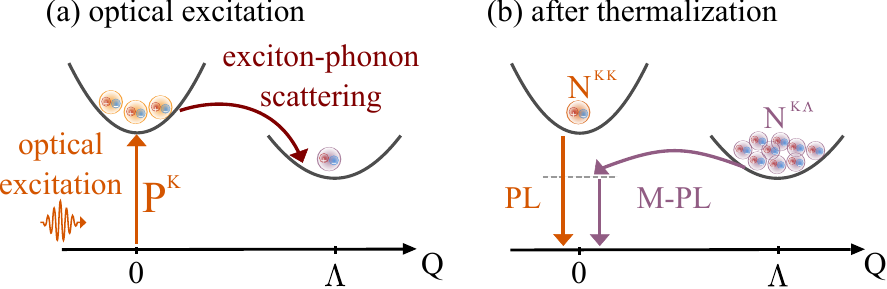} 
\end{center}
    \caption{\textbf{Schematic illustration of molecule-induced photoluminescence.} Excitonic dispersion of tungsten based TMDs with the K valley located at the center-of-mass momentum $Q=0$ and energetically lower $\Lambda$ valley located at $Q=\Lambda \approx  6.6 \,\text{nm}{^{-1}} $. (a) Optical excitation induces a microscopic polarization $P^{K}$ in the K valley. Due to exciton-phonon interaction the polarization can decay either within the K valley or to the energetically lower $\Lambda$ valley. At the same time incoherent excitons $N^{KK}$ and $N^{K\Lambda}$ are formed and thermalize until a Bose distribution is reached. (b)
 The bright excitons $N^{KK}$ located at the K valley decay radiatively by emitting a photon (PL), whereas the dark excitons $N^{K\Lambda}$ at the $\Lambda$ valley require a  center-of-mass momentum to decay back to the light cone and emit light. Molecules on the surface of the TMD material can provide this momentum and hence induce photoluminescence (M-PL) from the dark $\Lambda$ valley.   }
   \label{schema}
\end{figure}

\section{Theoretical approach}

\subsection{Photoluminescence}
To get a microscopic access to the optical response of pristine and molecule-functionalized monolayer TMDs after excitation with a laser pulse, we apply the density matrix formalism in combination with the nearest-neighbor tight-binding approach \cite{Kochbuch, Kira2006,carbonbuch, kadi14}.
Our goal is to calculate the steady-state photoluminescence  $\PL(\omega_q)$ which is given by the rate of emitted photons
\begin{equation}\label{SteadyStatePL}
\PL(\omega_q) \propto \omega_q \frac{\partial}{\partial t} \langle c^{\dagger}_{\bf q}c_{ \bf q}\rangle
 \propto \text{Im} \left [ 
\sum_{\bf{k_1}\bf{k_2}\mu}  M_{\bf q \bf{k_1}\bf{k_2}}
S^{vc_{\mu}}_{\bf{k_1}\bf{k_2}} (\omega_q )\right ]
\end{equation}
which is determined by the photon-assisted polarization \cite{thranhardt2000quantum}
  $S^{vc_{\mu}}_{\bf{k_1}\bf{k_2}} =  \langle c^{\dagger}_{\bf q}
a^{\dagger v}_{\bf{k_1}} a^{c_\mu}_{\bf{k_2}}
\rangle $ with electron annihilation (creation) $a^{(\dagger)}$ and photon annihilation (creation) $c^{(\dagger)}$ operators.
This microscopic quantity is a measure for emitting photons with the energy $\hbar \omega_q$ due to relaxation from the state ($c_{ \mu}, \bf{k_2}$) in the conduction band of valley $\mu$ with the electronic momentum $\bf k_2$ to the state ($v, \bf{k_1}$) in the valence band with the momentum $\bf k_1$. Note that we take into account the conduction band minima both at the K and the $\Lambda$ valley, while there is only a valence band maximum at the K valley. We neglect the impact of the energetically lower $\Gamma$ valley.

Before we derive the TMD Bloch equations, we account for the crucial importance of excitonic effects \cite{Chernikov2014, gunnar_prb,arora2015excitonic} by  transforming the system to the excitonic basis. We use the relation
$X^{cv}_{\bf{k_1 k_2}} \rightarrow
X_{\bf{qQ}}^{cv }= \sum_{\mu} \varphi_{\bf q}^{\mu} X_{\bf{Q}}^{\mu}$, where each observable $X_{\bf{qQ}}^{cv }$ is projected to a new excitonic quantity $X_{\bf{Q}}^{\mu}$ that is weighted by the excitonic wave function $\varphi_{\bf q}^{\mu}$.
For higher correlation it reads accordingly: 
$X^{cvvc}_{\bf{k_1 k_2 k_3 k_4}} \rightarrow
X_{\bf{qQq'Q'}}^{cvvc }= \sum_{\mu\mu'} \varphi_{\bf q}^{\mu} \varphi_{\bf q'}^{\mu' *} X_{\bf{QQ'}}^{\mu\mu'}$
Here, we have introduced the center-of-mass momentum $\bf Q = k_2 - k_1$ and the relative momentum ${\bf q}=\alpha {\bf k_1} + \beta {\bf k_2}$ with  $\alpha= \frac{m_h}{m_h + m_e^{\mu}}$ and $\beta=\frac{m_e^{\mu}}{m_h + m_e^{\mu}}$  with the electron (hole) mass $m_{e(h)}^{\mu}$.
 The excitonic eigenfunctions $\varphi_{q}$ and eigenenergies $\varepsilon^{\mu}$ are obtained by solving  the Wannier equation, which presents an eigenvalue problem for excitons \cite{Kochbuch, Kira2006}
\begin{equation}\label{wannier}
 \frac{\hbar^2 q^{2}}{2m^{\mu}}  \varphi_{\bf q}^{\mu} 
 - \sum_{\bf k} V_{\text{exc}}(\bf k)  \varphi_{\bf {q-k} }^{\mu}=\varepsilon^{\mu}\varphi_{\bf {q}}^{\mu}.
\end{equation}
Here, $m^{\mu}=\frac{m_h\cdot m_e^{\mu}}{m_h+m_e^{\mu}}$ is the reduced exciton mass and $V_{\text{exc}}$ describes the attractive part of the Coulomb interaction that is responsible for the formation of excitons. 
The corresponding Coulomb matrix elements are calculated within the Keldysh potential \cite{Keldysh1978, gunnar_prb,Cudazzo2011}.

For the case of pristine monolayer TMD one obtains the well-known Elliott formula for photoluminescence \cite{hoyer2005many} by solving the semiconductor luminescence equations \cite{kira1999quantum,thranhardt2000quantum}
\begin{equation}\label{PLG}
 I^{\sigma}(\omega_q) \propto
 \text{Im} \left[
\sum_{\mu} \frac{ 
|M^{\sigma\mu}|^2 \delta_{\mu,K}   \left( |P^{\mu}_0|^2 +  N^{\mu}_0\right) }
{\varepsilon^{\mu} - \hbar \omega_q - i\gamma^{\mu}}
\right].
\end{equation}
Here, $M^{\sigma\mu}$ is the exciton-photon matrix element corresponding to the coupling of the exciton to the $\sigma$ polarized light \cite{Kochbuch,thranhardt2000quantum}.
In contrast to absorption that is only determined  by the microscopic polarization
$P^{\mu}_{\bf Q}=
\sum_{\bf q}  
 \varphi_{\bf q}^{\mu*} 
\langle
a^{\dagger c_{ \mu}}_{{\bf q} +\alpha {\bf Q} } a^{v}_{{\bf q} -\beta {\bf Q} }
\rangle
$
the PL also shows an incoherent contribution that scales with the exciton occupation
$N^{\mu}_{{\bf Q}}=
\sum_{{\bf q_1} {\bf q_2}}
 \varphi_{\bf q_1}^{\mu*} 
 \varphi_{\bf q_2}^{\mu} 
\delta
 \langle
a^{\dagger c_{\mu}}_{\bf{q_1}+\alpha {\bf Q}} a^{ v}_{\bf{q_1}-\beta {\bf Q}}
a^{\dagger v}_{\bf{q_2}-\beta {\bf Q}} a^{c_{\mu}}_{\bf{q_2}+\alpha {\bf Q}}
\rangle$.
 This quantity  corresponds to the expectation value of correlated electron-hole pairs and is referred to as incoherent excitons \cite{thranhardt2000quantum}. The latter cannot be created through optical excitation with a coherent laser pulse, but are formed assisted by e.g. exciton-phonon scattering \cite{selig2017dark,thranhardt2000quantum}.
 Finally, the denominator of Eq. (\ref{PLG}) contains the excitonic eigenvalues $\varepsilon^{\mu}$ determining the  position of excitonic resonance in PL as well as the  the dephasing rate $\gamma^{\mu}$ responsible for the linewidth of excitonic transitions. The latter will turn out to be crucial for the visibility of dark excitonic states. Therefore, we microscopically calculate $\gamma^{\mu}$ including its radiative and non-radiative part, which is dominated by exciton-phonon scattering in the considered low-excitation limit. We find $\gamma^{KK}= 5$ (10) meV and  $\gamma^{K\Lambda}=1$ (8) meV for 77 (300) K \cite{selig2016excitonic} in WS$_2$.

Our goal is now to calculate the molecule induced changes in the photoluminescence, i.e. to what extent \eqref{PLG} changes in presence of molecules. 
In order to calculate molecule-induced photoluminescence, we start with the relation for the luminescence in general, i.e. \eqref{SteadyStatePL}, and derive the TMD Bloch equations for our system. Beside the photon-assisted polarization $S^{\mu}$, which is the key quantity for the steady state luminescence, we will also investigate the molecule-induced changes in exciton polarization $P^{\mu}$ and incoherent exciton densities $N^{\mu}$ appearing in  \eqref{PLG}. In the following, we refer to $P^{\mu}$ as the quantity describing coherent excitons.

To derive the TMD Bloch equations for the microscopic quantities $X=S,P,N$, we exploit the Heisenberg equation of motion $i\hbar  \dot{X} = [X,H]$.
The Hamilton operator $H=H_0+H_{c-c}+H_{c-p}+H_{c-ph}+H_{c-m}$ describes many-particle interactions and includes the free carrier and phonon contribution $H_{0}$, the carrier-carrier interaction $H_{c-c}$, the carrier-photon interaction $H_{c-p}$,  the carrier-phonon interaction $H_{c-ph}$ and  the carrier-molecule interaction $H_{c-m}$.  A detailed description of the Hamilton operator and the appearing matrix elements can be found in our previous work \cite{selig2016excitonic, majaTechnikPaper, gunnar_prb}. 

The carrier-molecule interaction is considered to be an interaction between excitons in the TMD and the dipole moment induced by the attached molecules. The molecules disturb the translational symmetry in the TMD lattice and soften the momentum conservation in the system. Depending on the molecular distribution and coverage certain momenta are favored offering the possibility to address certain otherwise dark excitonic states. The exciton-molecule coupling elements read
\begin{equation}
\label{G}
G_{\bf{Qk}}^{\mu\nu}= \sum_{\bf{q}} 
(
\varphi_{\bf {q}}^{\mu*} g^{cc}_{\bf{q}_\alpha, \bf{q}_\alpha + \bf{k}} \varphi_{\bf{q+\beta k}}^{\nu} 
 - 
\varphi_{\bf {q}}^{\mu*} g^{vv}_{\bf{q}_\beta - \bf{k}, \bf{q}_\beta} \varphi_{\bf {q-\alpha k}}^{\nu})
\end{equation}
with $\bf{q}_\alpha=\bf{q}-\alpha\bf{Q}$ and $\bf{q}_\beta=\bf{q}+\beta\bf{Q}$. It corresponds to a sandwich between  
 the involved excitonic wave functions and the carrier-dipole coupling 
$g_{\bf{kk'}}^{\lambda\lambda'} = \langle \Psi_{\bf k}^{\lambda} ({\bf{r}}) | 
\sum_l \phi^{\bf d}_l({\bf{r}})
| \Psi_{\bf k'}^{\lambda'} (\bf{r})
\rangle
$ corresponding to the expectation value of the dipole potential 
$
\phi^{{\bf d}}_l({\bf{r}}) = \frac{1}{4\pi\epsilon_0} \frac{\bf d \cdot (r-R_l)}{|{\bf r-R_l}|^3} 
$ that is
formed by all attached molecules and the tight-binding wave functions $\Psi^\lambda_{\bf k}(\bf r)$. 
As a result, the strength of the exciton-molecule  interaction is given by the molecular dipole moment $\bf d$ and the distance $\bf R_l$ of the molecules from the TMD surface  as well as by the number of attached molecules $l$ which can be translated to a molecular coverage $n_m$. Assuming homogeneously distributed molecules in x-y direction, one can write ${\bf R_l} = (x\cdot \Delta R, y\cdot \Delta R, R_z)$ with $x,y \in N $ and $\Delta R$ as an average distance between the molecules.
With this we find for the carrier-dipole coupling elements
\begin{eqnarray}\label{gg}
g_{\bf{k_{1}k_{2}}}^{\lambda_1 \lambda_2} &=& \frac{e_{0}}{2\pi\varepsilon_{0}\hbar} \sum_{x} n_m \delta_{| {\bf{k_1-k_2}}|, \frac{2\pi x}{\Delta R}} \sum_{j}C_{j}^{\lambda_1 *} ({\bf k_{1}}) C_{j}^{\lambda_2} ({\bf k_{2}}) \notag \\
&\times&   \delta_{\bf{k_{1}-k_{2}},\bf{q}} 
 \int d{\bf {q}}\frac{\bf d \cdot \bf{q}}{|\bf q|^{2}} e^{-R_{z}q_z} \label{molDetail}
\end{eqnarray}
with  the tight-binding coefficients $C^\lambda_j(\bf k)$, where $\lambda=v,c$ denotes the valence or the conduction band, while $j$ determines the contribution from different orbital functions \cite{gunnar_prb}. 

Applying the Heisenberg equation in excitonic basis, we obtain the following TMD Bloch equations 
\begin{eqnarray}
 \dot S_{\bf{kQ}}^{\mu}
&=&\hspace{-2pt}-i\Delta\tilde{\omega}_{\bf{kQ}}^{\mu}
S_{\bf{kQ}}^{\mu}
 \hspace{-2pt}+\hspace{-2pt}
M^{\sigma \mu}_{\bf{kQ}} \delta_{\bf{Q,0}} \hspace{-2pt}\left( |P_{\bf{0}}^{\mu}|^2 
\hspace{-2pt}+\hspace{-2pt} N_{\bf{Q}}^{\mu}\right) \notag \\
&+&  \sum_{\nu, \bf{Q'}}\hspace{-2pt}G_{\bf{QQ'}}^{\mu \nu}   S_{\bf{k, Q-Q'}}^\nu	  \label{SGleichung} \\
\dot{P}^{\mu}_\mathbf{Q}&=&-i\Delta {\omega}_{\bf{Q}}^\mu
 P^{\mu}_\mathbf{Q} 
+ \Omega^{\mu}_{{\bf Q}} 
  P^{\mu}_\mathbf{Q}  \delta_{{\bf Q, 0}}
+ \sum_{\nu, \bf{Q'}}
G_{\bf{QQ'}}^{\mu \nu}   P_{\bf{ Q-Q'}}^\nu	\label{PGleichung} \\
\dot{N}^{\mu}_\mathbf{Q}&=& 
\sum_{\nu, \bf{Q'}} \Gamma_{{\bf Q'}{\bf Q}}^{ \nu \mu , \text{in}} |P^{\nu}_{{\bf Q'}} |^2 \delta_{{\bf Q', 0}}
-\Gamma^{\mu}_{\text{rad}} N^{\mu}_{{\bf Q}} \delta_{{\bf Q, 0}} \notag \\
&+& 
 \sum_{\nu, \bf{Q'}} \left(
\Gamma_{{\bf Q'}{\bf Q}}^{ \nu \mu , \text{in}} N^{\nu}_{{\bf Q'}} 
-\Gamma_{{\bf Q}{\bf Q'}}^{  \mu \nu , \text{out}} N^{\mu}_{{\bf Q}} 
\right) \notag \\
&+& \sum_{\nu,{\bf Q'}} \hspace{-2pt} |G_{\bf{QQ'}}^{\mu \nu}|^2 \hspace{-2pt} \left(\hspace{-2pt} N_{\bf{ Q-Q'}}^\nu \hspace{-2pt} - \hspace{-2pt} N_{\bf{ Q}}^\mu\right)
\mathcal{L}_{\gamma_{\mu\nu}}\hspace{-2pt}(\varepsilon^{\nu}_{{\bf Q-Q'}}-\hspace{-2pt} \varepsilon^{\mu}_{{\bf Q}})  
 \label{NGleichung}
\end{eqnarray}
corresponding to a coupled system of differential equations for the photon-assisted polarization $S_{\bf{Q}}^{\mu}$, the microscopic polarization (coherent excitons) $P^{\mu}_\mathbf{Q}$, and the incoherent exciton occupation $N^{\mu}_\mathbf{Q}$.
 Here, we have introduced
 $\varepsilon^{\mu}_{{\bf Q}}=\varepsilon^{\mu}+\frac{\hbar^2 Q^{2}}{2M^\mu}$ with the total mass $M^{\mu}=m_h+m_e^{\mu}$
and 
  $\Delta  \omega_{\bf{Q}}^\mu= \frac{1}{\hbar} (\varepsilon^{\mu}_{{\bf Q}}
-i\gamma^\mu)$. In Eq. (\ref{SGleichung}), the transition frequency is additionally determined by the photon frequency $\omega_{\bf k}$ and reads 
$\Delta \tilde{\omega}_{\bf{kQ}}^\mu
=\Delta  \omega_{\bf{Q}}^\mu - \omega_{\bf k}$.
In Eq. \ref{NGleichung} $\mathcal{L}_{\gamma_{\mu\nu}}$ represents a Lorenzian function with width $\gamma^{\mu\nu}=\gamma_\mu+\gamma_\nu$.
Furthermore, $G_{\bf{QQ'}}^{\mu \nu}$ is the exciton-molecule matrix element, which enables molecule-induced coupling between different states $\mu$ and $\nu$ for all quantities $S,P,N$. 

Equations  (\ref{PGleichung}) and (\ref{NGleichung}) describe the optical excitation and decay of coherent excitons as well as the formation, thermalization, and decay of  incoherent excitons  (Fig. \ref{schema}(a)). In contrast,  \eqref{SGleichung} describes the radiative decay of the thermalized excitons including the molecule-assisted photoemission process (Fig. \ref{schema}(b)).  The coherent excitons are driven by the optical field $\Omega^{\mu}$ and decay radiatively and non-radiatively, which is both covered in the dephasing rate $\gamma^\mu$. 
The dephasing of coherent excitons leads to the formation of incoherent excitons, which is reflected by
the term $\propto |P^2|$ in \eqref{NGleichung}. The incoherent excitons can also decay radiatively with the rate $\Gamma^{\mu}_{\text{rad}}$, as long as they are located within the light cone with $\textbf Q\approx 0$. 
 Moreover, the incoherent excitons thermalize towards a thermal Bose distribution through exciton-phonon scattering. The corresponding out-scattering rate $\Gamma^{\mu\nu, \text{out}}_{{\bf{QQ'}}}$ describes processes from the state $(\mu, {\bf Q})$ to the state $(\nu, {\bf Q'})$, while the in-scattering rate $\Gamma^{\nu\mu , \text{in}}_{{\bf{Q'Q}}}$ describes the reverse process. 
More details on the scattering rates can be found in Ref. \onlinecite{selig2017dark}.

Since we are interested in the steady-state photoluminescence after exciton formation, we can decouple \eqref{PGleichung} and \eqref{NGleichung}  from \eqref{SGleichung}.  
We first need to solve \eqref{PGleichung} and \eqref{NGleichung} to get access to the thermalized exciton distribution. The results are presented in the next section, in particular focusing on the   changes in the exciton dynamics induced by the presence of molecules. 
With this, we have access to the steady-state photoluminescence by solving \eqref{SGleichung} via Fourier transformation. To get analytic insights, we can restrict the appearing sum over the momentum $\bf Q'$ in \eqref{SGleichung} by taking into account only the most pronounced terms with ${\bf Q'} = 0$. Hence we find for the incoherent contribution of the photoluminescence
\begin{equation}\label{PLana}
 I(\omega) \propto
 \text{Im} \bigg(
 \frac{|M_{\omega}^{\sigma K}|^2}{\Delta E_{\omega}^{K} - 
\frac{|G^{K\Lambda}|^2}{\Delta E_{\omega}^{\Lambda}}}
\left[
N_{\bf 0}^{KK} (1-\alpha)  
 + 
\alpha N_{\bf 0}^{K\Lambda} 
 \right]
\bigg)
\end{equation}
with $\alpha= \frac{|G^{K\Lambda}|^2}{(\Delta E_{\omega}^{\Lambda})(\varepsilon^{\Lambda}-\varepsilon^{K}+i\gamma^{K\Lambda})}$ and 
 $\Delta E_{\omega}^{\mu} = \varepsilon^{\mu} - \hbar \omega - i\gamma^{\mu}$. For $G^{K\Lambda}=0$, i.e. the pristine case without molecules, this leads to the well-known Elliott formula from \eqref{PLG} in the incoherent limit with $|P^{\mu}| = 0$. 
If $G^{K\Lambda}\neq 0$, i.e. in the case of molecules attached to the surface of the TMD,  we expect new peaks to appear in the PL, whenever $\Delta E_{\omega}^{\Lambda} = 0$.

Now, we have all ingredients at hand to investigate molecule-induced changes in the photoluminescence. If not otherwise stated, we use a standard set of molecular parameters: dipole moment of $d=13$ D corresponding to the exemplary merocyanine molecules \cite{photocrome}, a dipole orientation of \mbox{90 $^\circ$}, and  a molecular coverage $n_m=1.0 \text{ nm}^{-2}$. The orientation of the molecules is assumed to be perpendicular to the TMD plane, which is the most favorable case for densely packed molecules \cite{tsuboi2003formation}. Moreover, we assume the molecules to be attached non-covalently via van der Waals interaction leading to a distance between molecules and TMD surface of $R_z =0.36$ nm. We model the realistic situation, where TMD monolayers are located on a SiO$_2$ substrate with a dielectric constant of  $\epsilon_{\text{bg}} = 3.9$. We assume a typical carrier density of $n_{ex}=10^{11} \text{ cm}^{-2}$. To calculate the relative separation between bright $KK$ and dark $K\Lambda$ exciton, we solve the Wannier equation using consistent DFT input parameters regarding the electronic band structure of TMDs \cite{andor}. We find $\Delta E^{K\Lambda}= E^{KK}-E^{K\Lambda} \approx 50$ meV for WS$_2$ as our standard TMD material. Finally, we use an exemplary temperature of $T=77$ K as the linewidths in this regime are narrow enough to study the molecule-induced changes in the optical response. A detailed temperature study including optimal room temperature conditions is revealed in the last section of this manuscript.

First, we will show the influence of the molecules on the exciton dynamic by solving \eqref{PGleichung} and \eqref{NGleichung} in order to access the steady state exciton distribution needed for photoluminescence. With this insight, we will then calculate the molecule-assisted photoluminescence by solving \eqref{SGleichung} and exploiting \eqref{SteadyStatePL}.

\subsection{Exciton dynamics}\label{ExcDyn}

\begin{figure}[t]
  \begin{center}
\includegraphics[width=\linewidth]{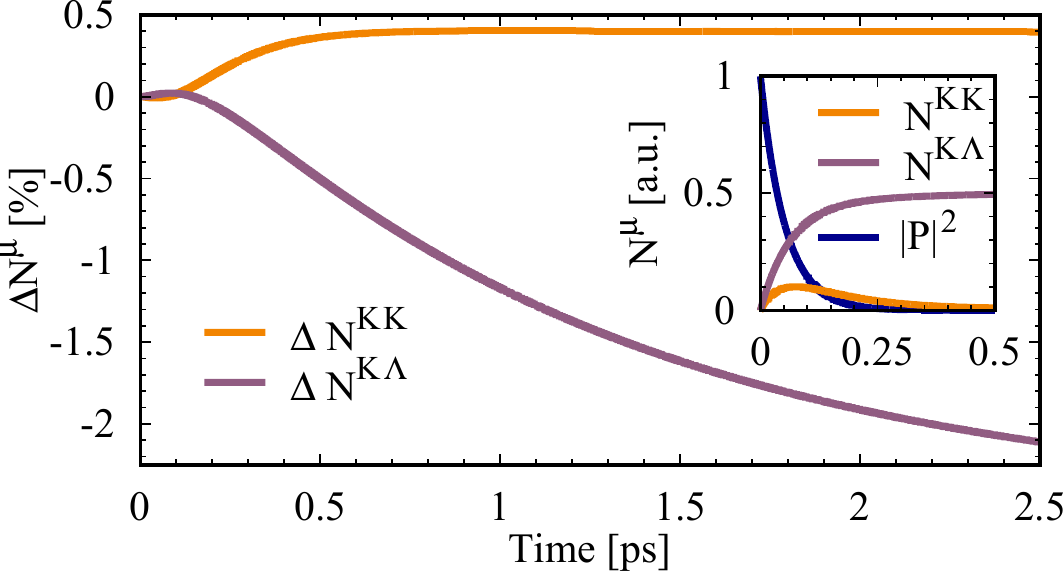} 
\end{center}
    \caption{\textbf{Impact of molecules on exciton dynamics.} 
Molecule induced changes $\Delta N^{\mu} = (N^{\mu}_{\text{mol}}-N^{\mu}_{0})/N^{tot}_0 $  
 in the occupation of $KK$ (orange line) and $K\Lambda$ (purple line) excitonic states in WS$_2$ functionalized with merocyanine molecules with a dipole moment of 13 Debye at the exemplary temperature of 77 K. Here, $N^{\mu}_{\text{mol}}$ and $N^{\mu}_{0} $ denote the excitonic occupation in and without the presence of molecules, respectively. We observe that the molecules first slightly enhance the formation of $K\Lambda$ excitons ($t<0.2$ ps), while during the exciton thermalization, they increase the population of  $KK$ excitons. 
In general, the molecule-induced changes in the exciton dynamics are rather small with less than  3$\%$. 
The inset shows the absolute exciton dynamics for pristine WS$_2$, where after 0.5 ps an equilibrium distribution is reached with the highest occupation of the energetically lowest $K\Lambda$ excitons. }
  \label{excDynamics}
\end{figure}

Before we investigate the changes in the optical fingerprint of the TMD material after non-covalent functionalization with molecules, we  first study the impact of molecules on the exciton dynamics.
Evaluating \eqref{PGleichung} and \eqref{NGleichung}, we have a microscopic access to the time- and momentum-resolved dynamics of coherent and incoherent exciton densities and can track the molecule-induced changes in the formation and thermalization of excitons. For pristine TMDs we find (i) the formation of a  coherent exciton density $|P^K|^2$ as the response to the optical excitation of the system with a weak pump pulse, (ii) 
radiative and non-radiative decay of the coherent exciton density and the phonon-assisted formation of an incoherent exciton density $N^{\mu}_{{\bf Q}}$, cf. the inset of Fig. \ref{excDynamics}. 
The timescale for the decay of coherent and the formation of incoherent excitons is rather fast with $<0.1$ ps after optical excitation, whereas the thermalization is on a slower timescale and the equilibrium is reached after approximately 0.5 ps \cite{selig2017dark}.
Interestingly, after thermalization the density of the $KK$ excitons is negligibly small, since most excitons occupy the energetically lowest $K\Lambda$ states (inset of Fig. \ref{excDynamics}). As this state is dark, i.e. optically inaccessible, most incoherent excitons are lost for optics. However, molecules can principally provide the required center-of-mass momentum $\bf Q$ to reach these dark states.

Now, we investigate how the attached molecules  influence the processes of exciton formation and thermalization. This is achieved by including the molecules in the calculation of coherent and incoherent exciton densities, cf. the last line in \eqref{PGleichung} and \eqref{NGleichung}. 
Figure  \ref{excDynamics} illustrates the difference of the densities $\Delta N^\mu = (N^\mu_{\text{mol}} - N^\mu_{0})/N^{tot}$ with and without the exemplary merocyanine molecules, normalized to the total occupation $N^{tot}$. 
We find that within the first \mbox{100 fs} the occupation of the $K\Lambda$ excitons is slightly enhanced, while the occupation of $KK$ excitons is reduced. This means that molecules support the formation of incoherent $K\Lambda$ excitons on the one hand  and suppress the formation of incoherent $KK$ excitons as molecule-mediated exciton relaxation to $K\Lambda$ states is very efficient. 
For the exciton thermalization the behavior for $K\Lambda$ and $KK$ excitons is inverse and the molecule-induced changes are more pronounced. 
Note however that the the observed changes are generally rather small and are in the range of $2 \%$. This justifies well the decoupling of  \eqref{PGleichung}, \eqref{NGleichung} from \eqref{SGleichung} as the influence from the molecules to the thermalized distributions is small. Furthermore, since the focus  of our work lies on the energy-resolved photoluminescence, it is sufficient to take into account thermalized incoherent exciton occupations. 

In the following, we investigate to what extent the molecule-induced photoluminescence is sensitive to experimentally accessible knobs, such as carrier excitation density and temperature as well as molecular characteristics including dipole moment, orientation, distribution, coverage, and distance. We focus on tungsten-based TMDs (WS$_2$ and WSe$_2$), since here the dark $K\Lambda$ exciton is the energetically lowest state exhibiting a large occupation after the thermalization. As a result, we expect the largest PL signal for dark excitons in W-based TMDs. 

\section{Excitation density}
Here, we study the impact of the excitation density on the PL of TMDs in presence of molecules.
For incoherent excitons after thermalization, we assume a Bose distribution
\begin{equation}\label{Bose}
 N^{\mu}_Q = \left[
\exp{\left(\frac{E_Q^{\mu} - \mu_{\text{chem}}}{k_{\text{B}} T}\right)} - 1
\right]^{-1}
\end{equation}
with $E_Q^{\mu} = \varepsilon^{\mu} + \frac{\hbar^2Q^2}{2M^{\mu}}$, the Boltzmann constant $k_{\text{B}}$ and the chemical potential \cite{Kochbuch}  $ 
\mu_{\text{chem}}
=
k_{\text{B}} T \ln \left[
1-\exp(-\frac{n_{\text{ex}}\hbar^2 2 \pi }{k_{\text{B}} T 3 M^{\mu}} )
\right]
$, where $n_{\text{ex}}$ is the excitation density. 
\begin{figure}[t]
  \begin{center}
\includegraphics[width=\linewidth]{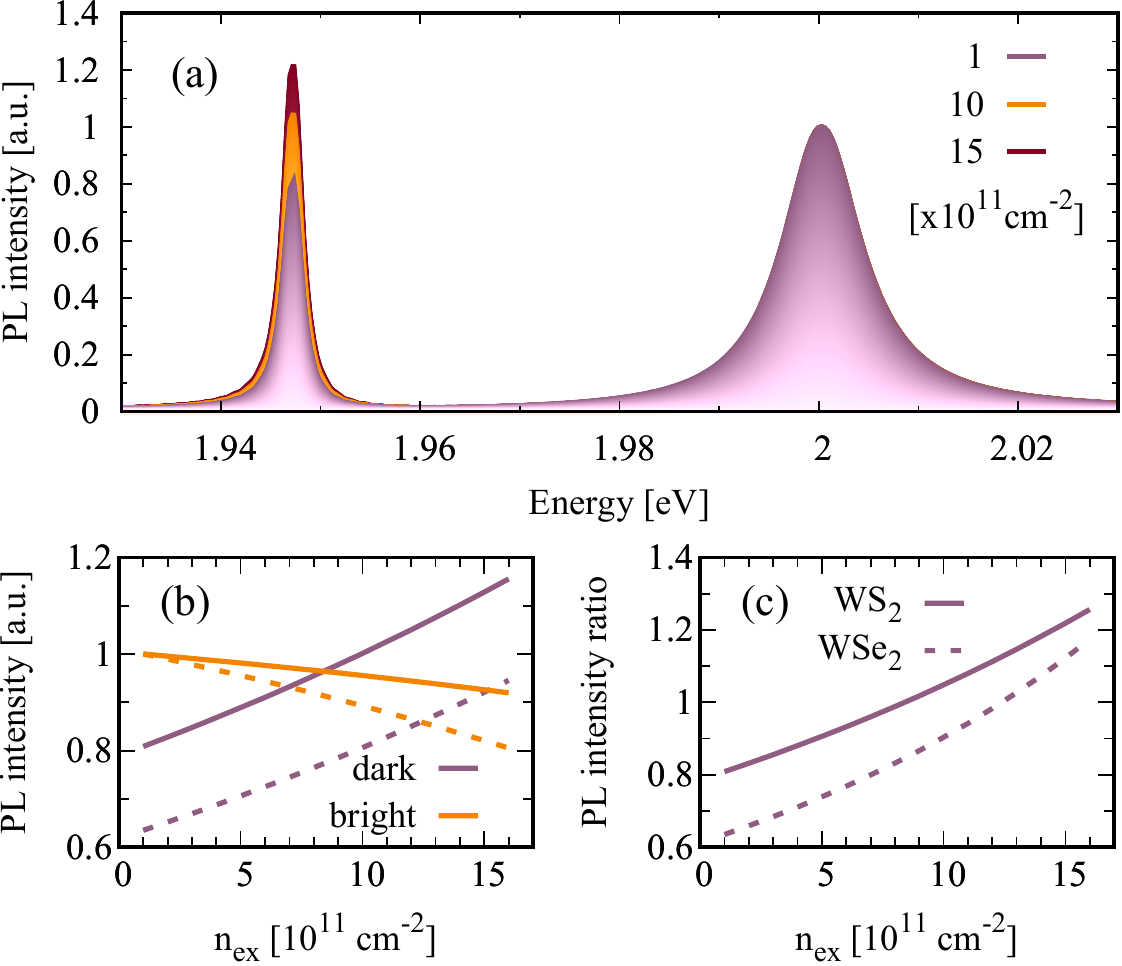} 
\end{center}
    \caption{\textbf{Dependence on excitation density.}
 (a) The PL spectrum of molecule-functionalized WS$_2$ shows a  broad peak at 2.0 eV stemming from bright $KK$ excitons and a narrow peak at  1.947 eV  reflecting the molecule-activated dark $K\Lambda$ excitons.
 The calculation is performed at 77 K and the PL intensity at different excitation densities is normalized to the the bright resonance. In the low excitation regime (purple curve) the dark peak reaches 80\% of the intensity of the bright peak. For  excitation densities higher than  $10 \times 10^{-11} \text{ cm}^{-2} $, the dark peak becomes even more pronounced.  (b) Maximum PL intensity of the dark (purple) and the bright (orange) peak as well as (c) their ratio as a function of the excitation density $n_{\text{ex}}$  for WS$_{2}$ (solid) and WSe$_{2}$ (dashed). Both TMDs show qualitatively the same behavior, although in WS$_{2}$ the dark peak is even for low excitation densities more pronounced due to larger overlap of excitonic wave functions resulting in more efficient molecule-exciton coupling.
}
  \label{excDensity}
\end{figure}
Now, we investigate the influence of $n_{\text{ex}}$  on $KK$ and $K\Lambda$ excitons and their optical fingerprint in PL spectra.  The incoherent exciton density is the driving mechanism for the photon-assisted polarization and hence for the PL, cf. \eqref{PLana}. For the pristine case, $N_{\bf{0}}^{KK}$ is the crucial quantity, while the relaxation from $K\Lambda$ excitons and hence the occupation $N_{\bf{0}}^{K\Lambda}$ only contributes in presence of molecules, i.e.  $G^{K\Lambda} \neq 0$.  Since the excitation density $n_{\text{ex}}$ directly enters  $N_{\bf{Q}}^{\mu}$, we expect it to have a large impact on the PL.

Figure \ref{excDensity}(a) shows the PL spectrum for the exemplary  monolayer WS$_2$ functionalized with merocyanine molecules. 
The broad peak located at 2.0 eV corresponds to photons emitted from the bright $KK$ excitons whereas the narrow peak   at 1.947 eV corresponds to molecule-induced emission of photons from dark $K\Lambda $ excitons, cf. also Fig. \ref{schema}(b) for an schematic view of the process. 
The linewidth of the energetically lower lying dark exciton is smaller, since it cannot decay radiatively and since also the non-radiative channels are restricted to less efficient processes involving the absorption of a phonon. 

As incoherent excitons are the driving force for photoluminescence, one can think about possibilities to optimize the signal from the dark exciton state via change of the excitation density $n_{ex}$. Figure \ref{excDensity}(a) shows the PL spectra for different $n_{ex}$. For a better comparison of the relative intensities, the spectra are normalized to the intensity of the bright peak. We observe that the dark exciton peak becomes more pronounced with increasing excitation density. Figure \ref{excDensity}(b) shows the absolute maximum intensity of the bright (orange line) and the dark (purple line) peak as a function of the excitation density. We find that the bright exciton  peak is decreasing, while the dark one is increasing in intensity. This can be ascribed to the fact that with higher excitation even more excitons occupy the energetically lower dark $K\Lambda$ state. 
Considering the PL intensity ratio of the dark to the bright exciton peak, we even find that for $n_{ex} >10 \cdot 10^{11}$ cm$^{-2}$ the dark peak becomes more pronounced than the bright one, cf. Fig. \ref{excDensity}(c). Even in the low excitation regime  ($n_{ex} < 10^{11} $ cm$^{-2}$ ) the dark exciton is still clearly visible in the PL. 

For comparison, we also show the PL for WSe$_2$ (dashed lines). We find a similar behavior. The qualitative differences stem from different exciton-molecule coupling strengths in the two materials and different masses $M^{\mu}$ entering the chemical potential. The exciton-molecule coupling is more efficient in WS$_2$ and hence the visibility of the dark peak is stronger in general. On the other hand, the higher mass $M^{\mu}$ in WS$_2$ reduces the sensitivity to the excitation density, i.e. the Bose distribution changes slower than in WSe$_2$ and hence WSe$_2$ shows a larger slope in the intensity ratio, cf. \ref{excDensity}(c).

To sum up, the excitation density is a promising experimental knob to enhance the visibility of the additional peak stemming from the molecule-activated  dark excitons. 

\section{Molecular characteristics}
Having revealed the principle mechanism of molecule-induced photoluminescence activating the dark $K\Lambda$ exciton, we now want to investigate the sensitivity of the mechanism on molecular characteristics. We have shown that the energetic position of the dark peak is determined by the internal dark-bright separation within the TMD,  whereas the intensity of the dark exciton peak is given by the strength of the coupling with the molecules.
In the following, we study the PL intensity ratio between the dark and the bright exciton peak, since this is the key quantity for efficiency of the activation of dark excitons and thus for the sensitivity of the molecule detection.

\subsection{Molecular dipole moment and orientation}

\begin{figure}[t]
  \begin{center}
\includegraphics[width=\linewidth]{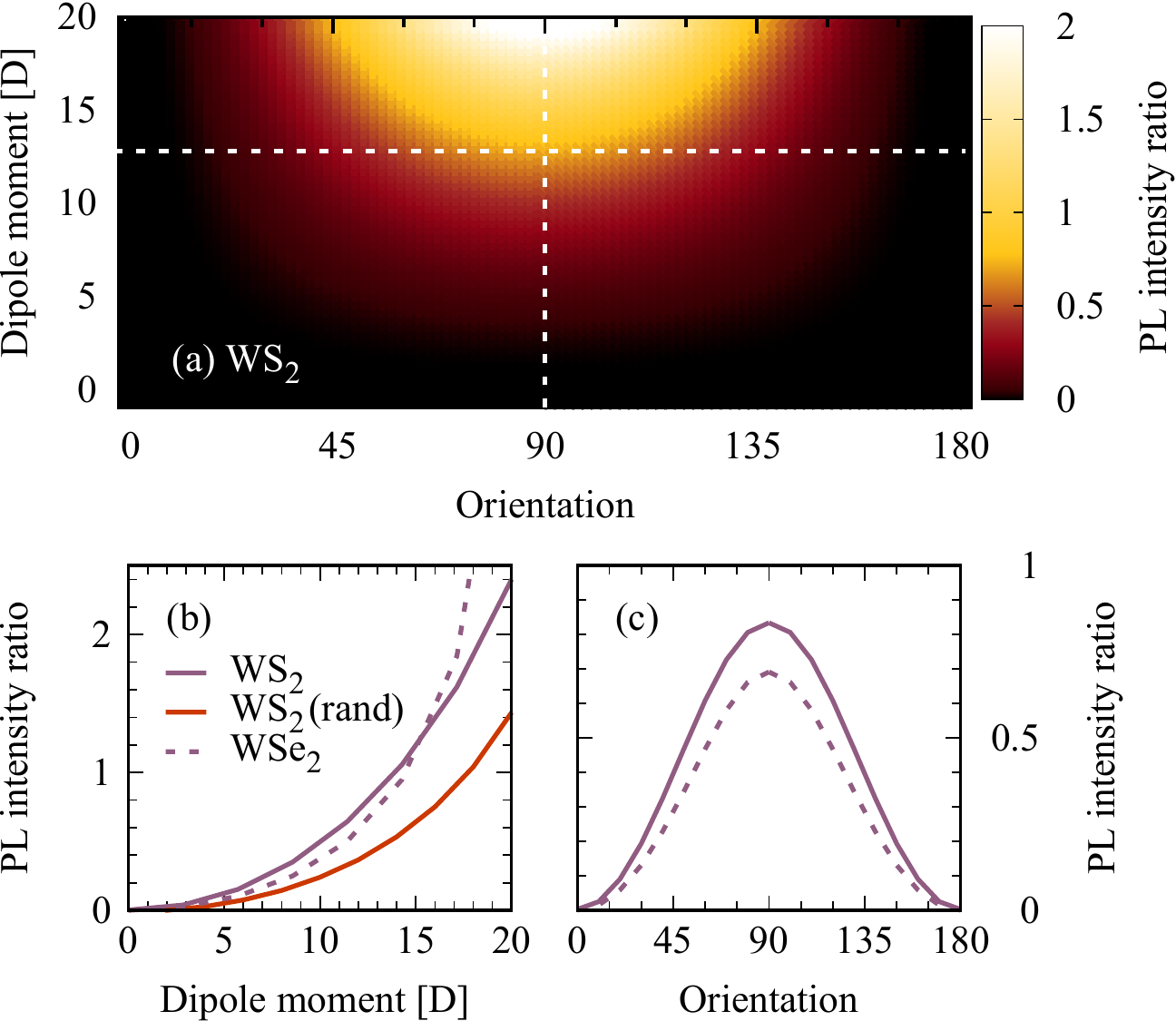} 
\end{center}
    \caption{\textbf{Dependence on molecular dipole moment and orientation.} (a) Surface plot showing the PL intensity ratio of dark and bright excitons in  functionalized WS$_{2}$ as a function of the molecular dipole moment and orientation at 77 K. We find the best visibility of the dark peak for perpendicular dipole orientation and large dipole moments. However, already for molecules with a dipole moment of 5 D, the dark exciton becomes visible in case of the perpendicular dipole orientation. The dashed white line shows our standard parameters within this manuscript including a fixed orientation of 90$^\circ$ and a fixed dipole moment of 13 D (merocyanine molecule). The corresponding cuts from the surface plot are shown in (b) and (c), respectively including a direct comparison to WSe$_{2}$  (dashed line). Additionally, we show the dependence on molecular dipole moment for randomized dipole orientation (orange line in (b)).
    }
  \label{MoleDipol}
\end{figure}

First, we study the impact of the molecular dipole moment including its orientation with respect to the TMD surface.  Figure \ref{MoleDipol}(a)  shows the PL intensity ratio between the dark and the bright exciton peak as a function of the strength and the orientation of the dipole moment in the case of functionalized WS$_2$ at 77 K. We find the best visibility of the dark exciton peak for high dipole moments with a perpendicular orientation. 

To obtain further insights, we show in Figs. \ref{MoleDipol}(b) and (c)  the dependence on the dipole moment for a fixed orientation of 90$^\circ$ and the dependence on dipole orientation for a fixed dipole moment of 13 D (merocyanine molecule), respectively (corresponding to the dashed white lines in Fig. \ref{MoleDipol}(a)). The  PL intensity ratio shows a quadratic increase with the dipole moment, and even for relatively low dipole moments of approximately 5 D the visibility of the dark exciton is in the range of 20$\%$. The dipole orientation study reveals a vanishing dark exciton feature for parallel orientation of the dipole moment and a maximum impact for perpendicular orientation.
Both observations can be understood in analogy to a classical dipole field. The stronger the dipole moment of the attached molecules, the stronger is the induced dipole field and hence the more efficient the molecules interact with the TMD. For the orientation of the dipole, the analogy to a classical dipole field reveals the largest overlap of the induced dipole field with the TMD surface for perpendicular dipole orientation.
The study on randomly orientated molecular dipole moments (cf. orange line in Fig.  \ref{MoleDipol}(b)) reveals that even though the dark-bright intensity ratio becomes smaller in case of randomization, we still obtain a visibility of $40 \%$ of the dark exciton peak  for d=13 D. 

Finally, the dashed lines in (b) and (c) show WSe$_{2}$ which reveals the same trends. The dark exciton peak for d$<$15 D is less pronounced than in WS$_{2}$ due to the less efficient exciton-molecule coupling. Interestingly, for stronger dipole moments (d$>$15D) the dark peak in WSe$_2$ becomes more visible. This can be traced back to the smaller total mass $M^{\mu}$ in WSe$_2$, which results in a more sensitive behavior to changes and eventually to a higher slope for the dipole dependence.

 In summary, we predict the best visibility of the dark exciton peak for molecules with a large dipole moment and a perpendicular orientation with respect to the TMD surface. A visibility of up to 10\% is predicted for molecules with a dipole moment of 3D.  

\subsection{Molecule-TMD distance}

\begin{figure}[t]
  \begin{center}
\includegraphics[width=\linewidth]{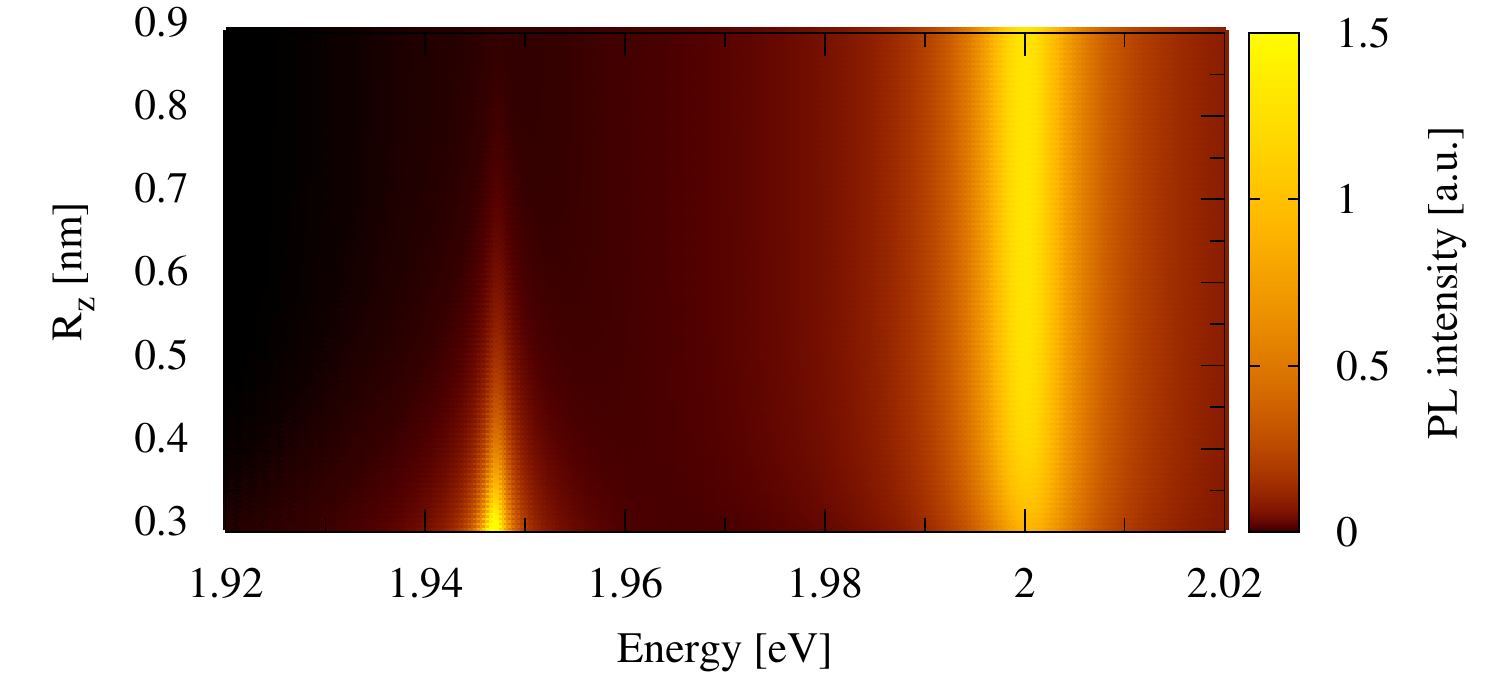} 
\end{center}
    \caption{\textbf{Dependence on molecular distance.} PL of functionalized  WS$_{2}$ in dependence of the distance between the TMD surface and the attached molecules $R_{z}$. The PL intensity is shown at 77 K and is normalized to the intensity of the bright peak. We find that the visibility of the dark exciton decreases with larger distance as the exciton-dipole interaction becomes weaker. The distance has no influence on the position of the dark peak. }
  \label{MoleDistance}
\end{figure}

Another crucial quantity is the distance $R_z$ between the molecules and the TMD surface. Here, we assume a non-covalent adsorption of molecules via van der Waals interaction. As a consequence, the minimal $R_z$ is given by the van der Waals radius which is approximately 0.36 nm. However, due to surface roughness, impurities, or the presence of linker molecules, this distance might be larger in a realistic experimental setup and hence it is important to shed light on the impact of  $R_z$ on the visibility of the dark exciton in PL spectra. 
Fig. \ref{MoleDistance} illustrates the maximum PL intensity as a function of energy and molecule-TMD distance  $R_z$. The bright exciton peak  is located at 2.0 eV and does not show any noticeable changes as a function of $R_z$. In contrast,  the visibility of the dark exciton PL peak at 1.947 eV is significantly reduced with the increasing distance. In analogy to the expectations from a classic dipole field, we find an exponential decrease of the exciton-molecule interaction with $e^{-R_z}$, cf. \eqref{molDetail}. 

In summary, the closer the molecules are attached to the TMD surface, the more pronounced is the  dark exciton feature in the PL spectrum. One could principally exploit the exponential dependence to determine the distance between the attached molecules and the material surface.

\subsection{Molecular distribution and coverage}

Now, we address the molecule distribution and coverage on the TMD surface and  to what extent they influence the visibility of the dark exciton in PL spectra.  The molecular coverage $n_m$ corresponding to the number of molecules on a fixed surface area plays a crucial role for the activation of dark excitons as it determines the induced center of mass momentum. Note that we do not consider molecule-molecule interactions, which might become important for a very large number of attached molecules.

Projecting the carrier-dipole coupling from \eqref{gg} into the excitonic basis and assuming the simplest case of a periodic molecular distribution allowing us to analytically solve the appearing integrals, we find 
\begin{equation}\label{Geinfach}
G_{Q} \propto \sum_{x}  \delta_{Q, \frac{2\pi x}{\Delta R}}  n_m e^{-Q}.
\end{equation}
One sees immediately the connection between the distance of molecules $\Delta R$ (reflecting the molecular coverage) in real space and the induced center-of-mass momentum in the Kronecker delta.
To reach the  $K\Lambda$ exciton,  a molecule-induced momentum transfer of approximately $Q\approx \unit[6.6]{nm^{-1}}$ is needed corresponding to the distance between the  $\Lambda$ to K valley in the Brillouin zone. This translates in  real space to $\Delta R=\frac{2\pi}{Q}\approx 1$ nm.  
For larger distances between the molecules (i.e. smaller molecular coverage), the momentum can still be provided through higher-order terms in the appearing sum in \eqref{Geinfach}, however the strength of the coupling becomes smaller. Note also that on one side the exciton-dipole coupling increases with larger molecular coverage, but on the other side it also decreases exponentially with the transferred momentum $Q$ (Eq. (\ref{Geinfach})). This results in an optimal molecular coverage, similarly to the already investigated case of carbon nanotubes \cite{malic11}.
 
\begin{figure}[t!]
  \begin{center}
\includegraphics[width=\linewidth]{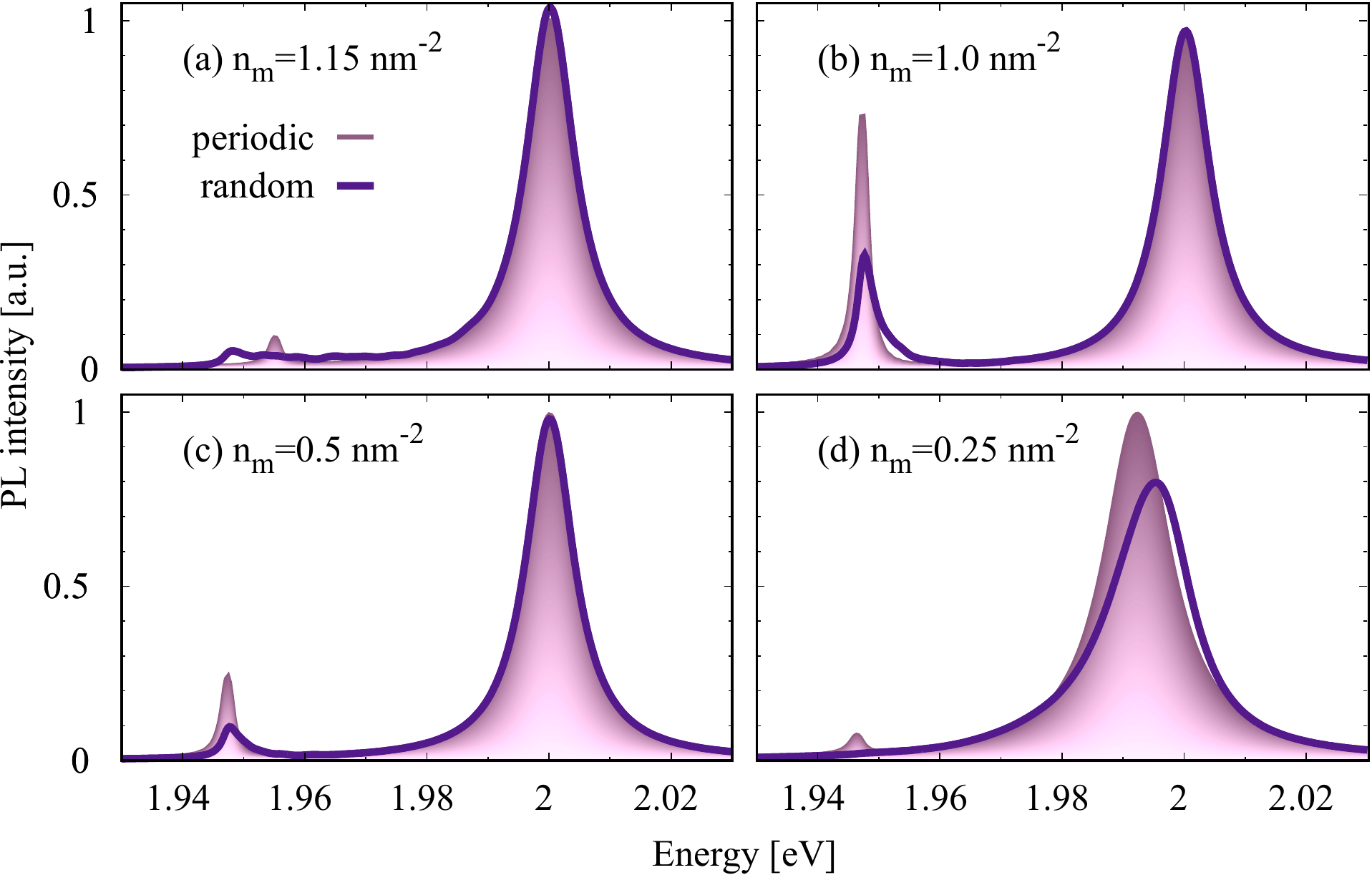} 
\end{center}
    \caption{\textbf{Dependence on molecular coverage.} PL spectra of functionalized WS$_{2}$  for a relatively  (a) high, (b) perfect, (c) medium, and (d) low molecule coverage $n_m$. The spectra are normalized to the bright peak. Light purple curves shows a periodic molecular distribution, whereas the dark purple line represents the case of randomly distributed molecules on the surface of WS$_2$. We find the best visibility for a molecular coverage of n=1.0 nm$^{-2}$ and a periodic distribution. In the case of randomly distributed molecules, the dark exciton peak decreases roughly to the half and smears out to the higher energy side. 
}
  \label{MoleCoverage}
\end{figure}

 First, we investigate the case, where molecules are periodically distributed on the TMD surface and build a molecular lattice. The corresponding molecular lattice constant determines the momentum that can be provided by the molecules to address dark excitonic states. Figure \ref{MoleCoverage} shows the PL spectra normalized to the bright peak for high ($n_m=1.15 \text{ nm}^{-2}$), the standard density in this manuscript ($n_m=1.0 \text{ nm}^{-2}$),  medium ($n_m=0.5 \text{ nm}^{-2}$), and low ($n_m=0.25 \text{ nm}^{-2}$) molecular coverage. Our calculations reveal that the visibility of the dark exciton is the most pronounced for $n_m=1.0 \text{ nm}^{-2}$ as the provided momentum $Q$ corresponds to the momentum needed to reach the $\Lambda$ valley. If we go to higher coverage, the peak decreases as the transferred momentum is not matching the $\Lambda$ valley and the coupling strength decreases exponentially with $Q$, cf. \eqref{Geinfach}.
Note however that even for small coverage, the dark exciton peak is still clearly visible. Its intensity is  in the range of  10$\%$ of the intensity of the bright peak. 

We observe that the bright peak shifts to the red at small molecular coverage. 
The origin of the red shift can be understood as follows: the transferred momentum for $n_m=0.25 \text{ nm}^{-2}$ of  $Q\approx 1.65 \text{ nm}^{-1}$ is rather small and can only enable  indirect transitions within the dispersion of the  $KK$ excitons with an energy $E^{KK}=\varepsilon^{K} + \frac{\hbar^{2}Q^{2}}{2M^{K}}$.
These intravalley transitions become more favorable for small molecular coverage. The molecule-induced PL from these states is approximately 100-200 meV above the bright peak (not shown in the spectra).  The coupling to the bright resonance leads to the observed red-shift. A detailed description of intravalley transitions can be found in Refs. \onlinecite{majaTechnikPaper, ermin_prl, gunnar_carbon}. The focus in this manuscript is on higher coverage, where the dark $K\Lambda$ can be reached. 

Now, we investigate the case of randomly distributed molecules on the TMD surface. We assume a fluctuation of the position of the molecules around their equilibrium position. We allow fluctuations of up to 10$\%$ modeled by a Gaussian random distribution. We find that now the dark exciton peak smears out to higher energies and its intensity decreases by approximately the half. The dark exciton remains visible, if the molecular coverage is not too low, cf. dark purple lines in Fig. \ref{MoleCoverage}. The random distribution of  molecules weakens the efficiency of exciton-molecule coupling and reduces the visibility of the dark exciton. For low molecular coverage, the dark exciton peak even disappears (Fig. \ref{MoleCoverage}(d)),  as the probability to find the required center-of-mass momentum to address the intervalley $K\Lambda$ excitons is low. However, the probability for intravalley  transitions along the dispersion of the $KK$ exciton increases. This $KK$ transitions are responsible for the red shift (still observable, but less pronounced compared to the periodic case due to the reduced exciton-molecule coupling). 
The observed peak asymmetry in case of the randomized distribution can be traced back to dark $K\Lambda$  transitions along the dispersion of the $K \Lambda $ exciton corresponding to $E^{\Lambda\Lambda}=\varepsilon^{\Lambda} + \frac{\hbar^{2}Q^{2}}{2M^{\Lambda}}$. Due to the large effective mass $M^{\Lambda}$ the molecule induced PL from these states is approx. 50 meV above the dark peak and hence it is visible as a high-energy wing. 

In summary, low molecular coverage and randomized molecular distributions induce peak broadening and asymmetry as well as a reduction of the dark exciton peak intensity. Nevertheless, the dark exciton still remains visible confirming that the effect is robust also under realistic conditions.  The best visibility is clearly reached for high molecular coverage and periodically distributed molecules.

\section{Temperature dependence}

\begin{figure}[t]
  \begin{center}
\includegraphics[width=\linewidth]{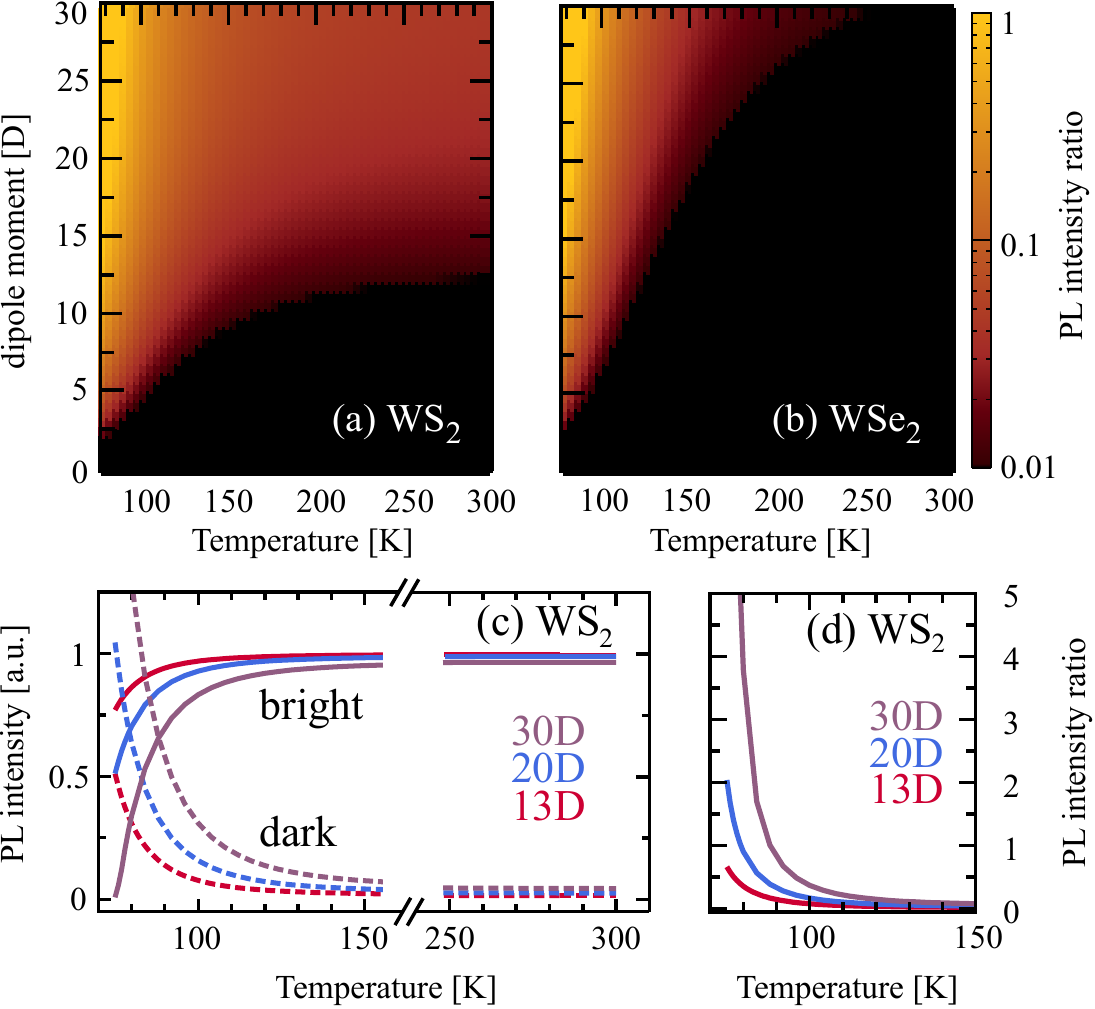} 
\end{center}
    \caption{\textbf{Dependence on temperature. }PL intensity ratio of dark-to-bright peak of functionalized (a) WS$_{2}$ and (b) WSe$_{2}$ for different temperatures $T$ and molecular dipole moments $d$. We find the best visibility of the dark exciton for low $T$ and high $d$.  Note that in case of WS$_{2}$ even at T=300 K the dark peak is visible for dipole moments down to 10 D. In case of WSe$_{2}$ the peak disappears at 250 K even for high dipole moment due to the broader linewidths in selenium-based TMDs. (c) PL intensity of the dark (dashed line) and bright (solid line) peak as a function of temperature for three exemplary dipole moments. We observe that, depending on the dipole moment, the crossing temperature between dark and bright peak changes. (d) PL intensity ratio of the dark and bright peak in WS$_{2}$ shows an exponential   decrease that can be ascribed to the Bose distributions of thermalized excitons.  }
  \label{TempiSurface}
\end{figure}

The investigations discussed so far have been performed at a temperature  of 77 K, since here the exciton-phonon coupling is weak leading to narrow linewidths allowing a clear separation of bright and dark exciton peaks in PL spectra. Now, we show  a temperature study on the visibility of the dark exciton aiming at the possibility for room temperature detection of molecules. 

The temperature affects both the  Bose-Einstein distribution of phonons and excitons. The first has a direct impact on  the efficiency of the exciton-phonon coupling and hence the exciton linewidths. The second determines the relative occupation of dark and bright exciton states directly influencing the PL spectra.
Figures \ref{TempiSurface}(a) and (b) illustrate the PL intensity ratio between the dark and the bright peak as a function of temperature and molecular  dipole moment for WS$_{2}$ and WSe$_{2}$, respectively. For low temperatures and high dipole moments, the visibility of the dark exciton is the best, as the peak linewidths are narrow and the exciton-dipole coupling is strong.
We find a much broader temperature and dipole moment range with a good visibility of the dark exciton in WS$_{2}$. 
  Surprisingly, we observe at room temperature and at dipole moments  down to 10 D a clearly visible dark exciton peak. In contrast, for WSe$_2$ dark excitons cannot be efficiently activated at room temperature even for very high dipole moments of  above 30 Debye. This is  due to the enhanced carrier-phonon coupling, which  leads to broader peaks in selenium-based TMDs \cite{selig2016excitonic, DominikPhonons}.

For a more quantitative understanding, we show for WS$_{2}$ the maximum PL intensity for the bright (solid lines) and the dark peak (dashed lines) as a function of temperature for  three exemplary dipole moments, cf.  Fig. \ref{TempiSurface}(c). We see that the bright peak increases in intensity, whereas the dark peak decreases for higher temperatures. This reflects the Bose-Einstein distribution of thermalized excitons, where the occupation of the energetically higher bright state becomes larger with temperature.
For our standard set of parameters including a  dipole moment of 13 Debye, the bright exciton peak is more pronounced at all temperatures. However,  for 20 D (30 D), the dark peak becomes more efficient at least at 77K and exceeds the bright transition  by a factor of 2 (5) reflecting the strong exciton-molecule interaction. We observe that in the temperature range of $80-100$ K the dark exciton peak decreases quickly for all  dipole moments, while the bright peak becomes much more pronounced.  The critical temperature at which the bright exciton becomes more pronounced than the dark one is 80 (90) K for 20 (30) D. For $T>100 K$, the intensity of the bright peak saturates to 1, while the dark peak basically vanishes. 
Figure \ref{TempiSurface}(d) shows the PL intensity ratio of the dark and the bright exciton for WS$_{2}$ in the low temperature range of $70-150$ K. We observe a clear exponential decrease with temperature reflecting the Bose-Einstein distribution of thermalized excitons. The decay rate is the same for all molecular dipole moments, since the exciton-molecule coupling only determines the initial value for the PL intensity ratio, but not its decay. It is given by the Bose-Einstein distribution and the energetic difference between bright and dark state.

The general temperature dependence can be understood on microscopic footing: 
At low temperatures, the majority of excitons occupies the energetically lowest $K\Lambda$ states, as the exciton-phonon scattering is too weak to scatter carriers into the higher $KK$ exciton states. The larger the temperature, the more efficient exciton-phonon scattering and the larger is the redistribution of excitons among the $K\Lambda$ and $KK$ states, i.e. $N^{K\Lambda}$ decreases and $N^{KK}$ increases. As a direct consequence, the PL intensity of bright $KK$ excitons is enhanced at higher temperatures, while the PL intensity of the dark $K\Lambda$ is reduced. Additionally, the larger the temperature, the broader are the linewidths of both bright and dark excitonic resonances eventually resulting in a vanishing visibility of the dark exciton. 

\begin{figure}[t]
  \begin{center}
\includegraphics[width=\linewidth]{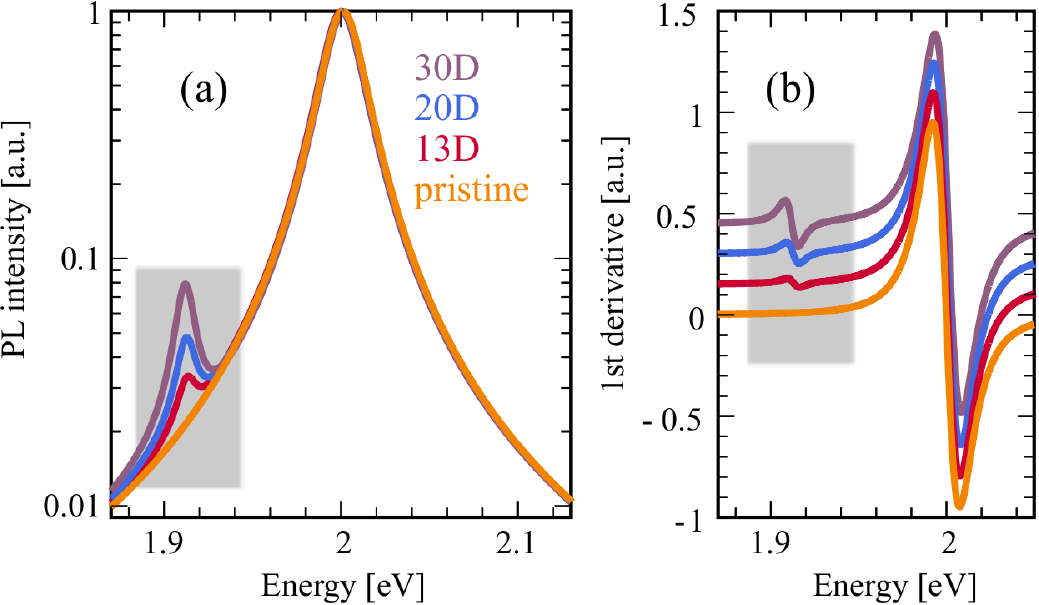} 
\end{center}
    \caption{\textbf{Room temperature conditions. } 
(a) Room temperature PL in logarithmic plot and (b) first derivative of the PL in functionalized and pristine WS$_{2}$ for different molecular dipole moments. The PL is again normalized to the bright peak. We find that the visibility of the dark peak is in the range of 3-9 $\%$ compared to the bright peak. (b) The main limiting factor is the broad excitonic linewidth at room temperature. Hence, we also show the derivative of the PL, which shows clear traces of the dark exciton at room temperature even for 13 D.  Note that the spectra are shifted along the y axes for better visibility. 
}
  \label{TempiSpektrum}
\end{figure}

As shown in Fig. \ref{TempiSurface}(a), the dark exciton peak is visible in WS$_{2}$ even at room temperature. To get further insights into room temperature conditions, we show the logarithmic PL spectrum at T=300 K for pristine (orange) and functionalized WS$_2$ including three different molecular dipole moments, cf.  Fig. \ref{TempiSpektrum}(a). 
In case of pristine WS$_{2}$, we find only one broad peak at 2.0 eV corresponding to the bright excitons resonance. With molecules, the additional peak at 1.91 eV  appears. Here, we investigate free-standing WS$_{2}$ without any substrate. This shifts the dark exciton peak to lower energies, which is favorable at room temperature conditions with large excitonic linewidths. Under these conditions, we find a visibility of the dark exciton peak to be approximately 3$\%$ of the bright exciton peak in the case of our standard merocyanine molecules with 13 D. For higher molecular dipole moments of 20 D (30 D), the intensity of the dark peak increases to 5 $\%$ (9 $\%$) with respect to the pristine peak, which should be resolvable in PL experiments. Even clearer signatures can be seen in  the first derivative of the PL spectrum, where we find an oscillation at 1.91 eV corresponding to the position of the dark $K\Lambda$ exciton.
This presents a large advantage compared to absorption spectra, where only a small  shoulder is visible at room temperature \cite{maja_sensor}. 

\section{Discussion and Conclusion}
Here, we summarize and discuss the obtained insights for low and room temperature conditions.
In the low temperature (77 K) case,  exciton-phonon coupling is relatively weak and PL spectra are characterized by narrow excitonic linewidths. These are good conditions for pronounced features stemming from the dark exciton.  We find that molecules with dipole moments larger than 3 Debye can be detected. The larger the dipole moment, the more efficient is the exciton-molecule coupling and the more pronounced is the dark exciton. Furthermore, the orientation of the dipole moment also matters. We find the largest visibility of the dark exciton for the perpendicular orientation with respect to the TMD surface, since here the overlap with the dipole field is the largest. Another important property is the molecular coverage $n_M$, since it determines the possible molecule-induced momentum transfer. We find an optimal response for $n_M\approx 1 \text{ nm}^{-2}$ in the case of periodically distributed molecules.  For randomized distributions, the effect becomes smaller, but the signatures of the dark exciton are still visible. 
Moreover, the smaller the distance of the attached molecules to the TMD surface, the more pronounced is the effect. We have shown that even for distances larger than the van der Waals radius, the molecule signatures still remain visible in  PL spectra.

In the room temperature case, the exciton-phonon coupling is strong  resulting in a broadening of the excitonic transitions, which strongly restricts the visibility of the dark exciton. However, for molecules with a dipole moment larger than  10 Debye, we can still observe clear signatures in the PL spectra assuming that the dipole orientation, the distance to the TMD surface, and the molecular coverage are optimal. Moreover, the excitation density can be used as an additional knob to further increase the visibility of the dark peak. 

In summary, we have revealed a promising potential of monolayer tungsten disulfide (WS$_2$) as a novel  nanomaterial for detection of molecules with a large dipole moment. We have shown that its photoluminescence is very sensitive to external molecules giving rise  to a well pronounced additional peak that can be ascribed to the activation of dark excitonic states. Depending on different experimentally accessible knobs, even room temperature detection of molecules becomes possible.

\section{Acknowledgement}
This project has received funding from the European Union's Horizon 2020 research and innovation programme under grant agreement No 696656 within the Graphene Flagship and the Swedish Research Council. Furthermore, we acknowledge support by  the Chalmers Area of Advance in Nanoscience and Nanotechnology. M.S. acknowledges financial support from the Deutsche Forschungsgemeinschaft (DFG) through SFB 787.


%

\end{document}